\documentclass[prx,twocolumn,floatix,amssymb,showpacs,amsmath, superscriptaddress,longbibliography]{revtex4-1}
\usepackage{graphicx}
\usepackage{upgreek}
\usepackage{epsfig,color}
\usepackage{amsmath,amssymb,amsthm}
\usepackage[utf8]{inputenc}
\usepackage[colorlinks, urlcolor=blue, linkcolor=red]{hyperref}

\usepackage{soul}
\usepackage{ dsfont }
\usepackage{tikz}
\usepackage[normalem]{ulem}

\newcommand{\beqa}{\begin{eqnarray}}
\newcommand{\eeqa}{\end{eqnarray}}

\renewcommand{\boxed}[2]{\textcolor{#1}{%
\tikz[baseline={([yshift=-1ex]current bounding box.center)}] \node [rectangle, minimum width=1ex,rounded corners,draw] {\normalcolor\m@th$\displaystyle#2$};}}

\usepackage{bookmark}

\begin{document}

\title{Quantum Illumination with a Hetero-Homodyne Receiver and Sequential Detection}
\author{Maximilian Reichert}
\email{maximilian.reichert@ehu.eus}
\affiliation{Department of Physical Chemistry, University of the Basque Country UPV/EHU, Apartado 644, 48080 Bilbao, Spain}
\affiliation{EHU Quantum Center, University of the Basque Country UPV/EHU, Bilbao, Spain}
\author{Quntao Zhuang}
\email{qzhuang@usc.edu}
\affiliation{Ming Hsieh Department of Electrical and Computer Engineering, University of Southern California, Los Angeles, California 90089, USA}
\affiliation{Department of Physics and Astronomy, University of Southern California, Los Angeles, California 90089, USA}
\author{Jeffrey H. Shapiro}
\email{jhs@mit.edu}
\affiliation{Research Laboratory of Electronics, Massachusetts Institute of Technology, Cambridge, Massachusetts 02139, USA}
\author{Roberto Di Candia}
\email{roberto.dicandia@aalto.fi}
\affiliation{Department of Information and Communications Engineering, Aalto University, Espoo, 02150 Finland}
\affiliation{Dipartimento di Fisica, Universit\`a degli Studi di Pavia, Via Agostino Bassi 6, I-27100, Pavia, Italy}
\date{\today}

\begin{abstract}

We propose a hetero-homodyne receiver for quantum illumination (QI) target detection.  Unlike prior QI receivers, it uses a cascaded positive operator-valued measurement (POVM) that does not require a quantum interaction between QI's returned radiation and its stored idler.  When used without sequential detection its performance matches the 3\,dB quantum advantage over optimum classical illumination (CI) that Guha and Erkmen's~[Phys. Rev. A {\bf 80}, 052310 (2009)] phase-conjugate and parametric amplifier receivers enjoy.  When used in a sequential detection QI protocol, the hetero-homodyne receiver offers a 9\,dB quantum advantage over a conventional CI radar, and a 3\,dB advantage over a CI radar with sequential detection. Our work is a significant step forward toward a practical quantum radar for the microwave region, and, more generally, emphasizes the potential offered by cascaded POVMs for quantum radar. 

\end{abstract}

\maketitle

\section{Introduction \label{Introduction}} Quantum radars use resources unavailable to their classical counterparts, principally entanglement, to obtain improved remote-sensing performance at the same transmitted energy, see Refs.~\cite{Shapiro2020,Torrome2020,Sorelli2022} for recent reviews.  To date, the only quantum radar protocol whose target-detection performance is predicted to exceed that of its best classical competitor is Tan~\emph{et al}.'s quantum illumination (QI)~\cite{Tan2008}.  QI with optimum reception offers a 6\,dB quantum advantage in error-probability exponent for detecting a weakly-reflecting target embedded in high-brightness (many photons/s-Hz) background noise.  This advantage \emph{only} occurs in a lossy, noisy setting that destroys the initial entanglement between QI's transmitted signal and its stored idler. In particular, Nair~\cite{Nair2011} has shown that in the absence of noise conventional coherent-state radar closely approximates the target-detection performance of the optimum quantum radar of the same transmitted energy. So, because daytime background light at near-visible wavelengths has extremely low brightness, e.g., $10^{-6}\,$photon/s-Hz at 1.55\,$\upmu$m wavelength~\cite{Shapiro2005}, Tan~\emph{et al.}'s QI attracted little interest from the radar community until Barzanjeh~\emph{et al.}~\cite{Barzanjeh2015} described how it might be used at microwave wavelengths, where high-brightness background noise is the norm and QI's quantum advantage could help in detecting stealth targets.

Tan~\emph{et al.}'s QI relies on the nonclassical phase-sensitive cross correlation between the brightness-$N_S$ signal and idler beams produced by a spontaneous parametric downconverter (SPDC), viz., signal and idler consisting of $M \gg 1$ independent and identically-distributed (iid) mode pairs in two-mode squeezed-vacuum (TMSV) states. The TMSV's nonclassical cross correlation, $\sqrt{N_S(N_S+1)}$, greatly exceeds the classical limit, $N_S$, in low-brightness ($N_S \ll 1$) operation regime, and disappears as $N_S$ grows without bound.  Furthermore, because conventional interference techniques are incapable of detecting phase-sensitive correlation~\cite{Shapiro2020}, the first proposed receivers~\cite{Guha2009} for obtaining \emph{any} quantum advantage from QI used parametric amplifiers to convert phase-sensitive correlation into phase-insensitive correlation prior to detection by conventional techniques.  These proposals---Guha and Erkmen's parametric amplifier (PA) and the phase-conjugate (PC) receivers---deliver \emph{at most} a 3\,dB quantum advantage in error-probability exponent, and to do so they require a quantum memory capable of losslessly storing the idler's high time-bandwidth product quantum state for the roundtrip radar-to-target-to-radar propagation delay.  So far, however, only 20\% quantum advantage has been demonstrated in optical wavelength (with high-brightness noise injection)~\cite{Zhang2015}  and microwave wavelength~\cite{Assouly2022} table-top experiments.  

The first explicit architecture for obtaining QI's full 6\,dB quantum advantage was the feed-forward sum-frequency generation receiver~\cite{Zhuang2017}, whose implementation requires an as yet unavailable single-photon nonlinearity as well as a quantum memory for idler storage.  A more recent architecture, the correlation-to-displacement receiver~\cite{Shi2022}, circumvents the need for a single-photon nonlinearity, but requires a lossless $M\times M$ programmable beam splitter with $M\gg 1$---which will be a daunting implementation task at microwave wavelengths---as well as the aforementioned quantum memory for idler storage.  Were available technology capable of realizing such QI receivers, the ultimate performance for Tan~\emph{et al.}'s target-detection scenario would be obtained, because theory~\cite{DePalma2018,Nair2020,DiCandia2021,Bradshaw2021, Sanz2017, Gong2022,Jonsson2022} has proven the optimality of the $M$ mode-pair TMSV state for that setting.

\begin{figure}[t!]
\centering
{\includegraphics[width=0.87\columnwidth]{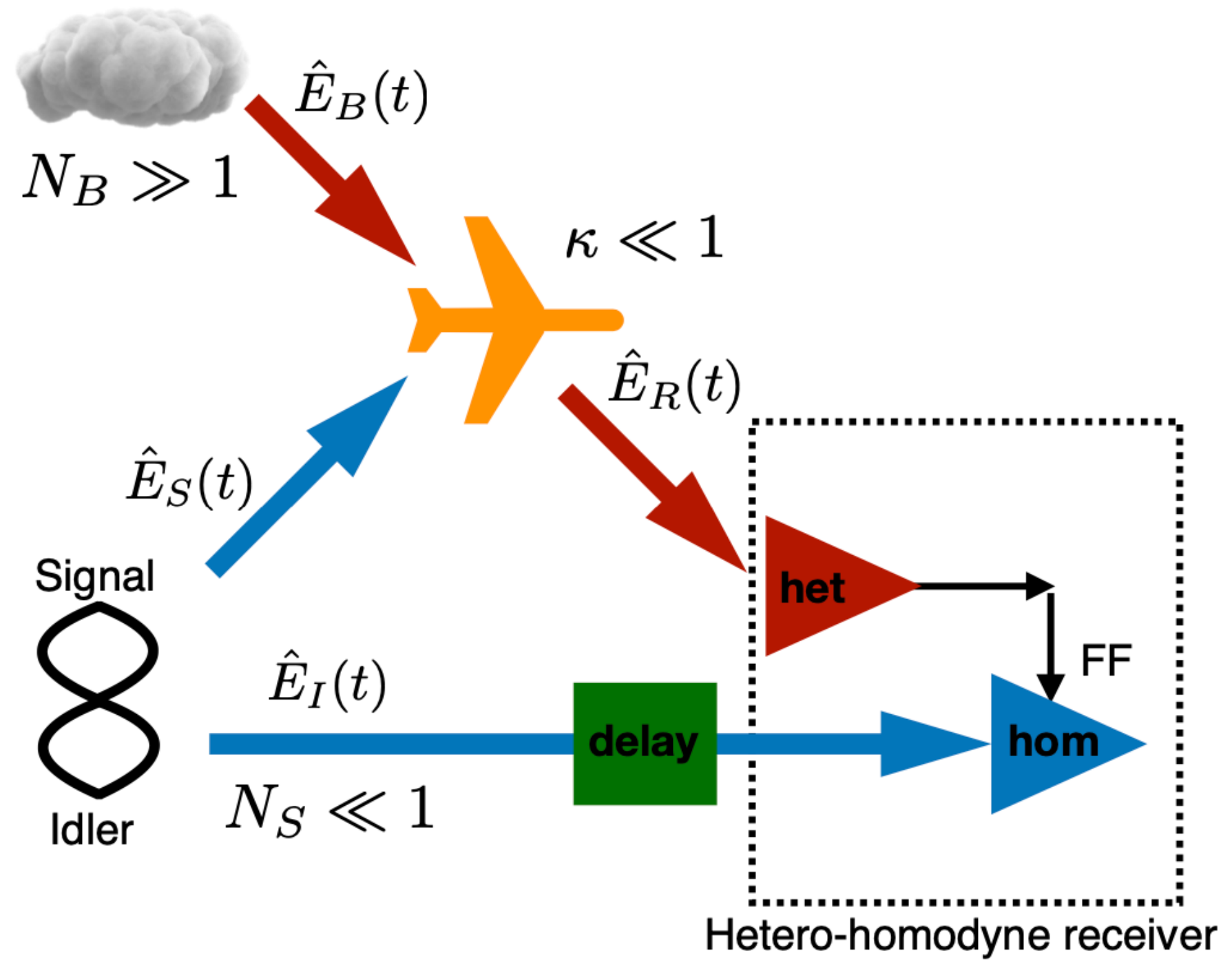}}
\caption{\label{Fig_QI}Sketch of the Tan~\emph{et al}.’s QI protocol with hetero-homodyne reception. A signal-idler system is initialized in an $M$ mode-pair TMSV state, with each signal and idler mode containing $N_S \ll 1$ photons on average. The signal is sent to test for the presence of  a weakly-reflecting (roundtrip transmissivity $\kappa\ll1$) target embedded in high-brightness background noise ($N_B\gg1$\, photons/s-Hz). The hetero-homodyne receiver measures the cross correlation between the returned radiation and the idler. Because the  low-brightness TMSV mode pairs’ phase-sensitive cross correlation $\sqrt{N_S(N_S+1)}$ greatly exceeds the classical limit $N_S$, QI outperforms classical illumination in this regime even though loss and noise have destroyed the TMSV states’ initial entanglement. het: heterodyne. hom: homodyne. FF: feed-forward.}
\end{figure}

This paper reports a significant advance for microwave QI, and, more generally, emphasizes the potential offered by cascaded positive-operator valued (POVM) measurements for quantum radar.  First, motivated by Shi~\emph{et al}.~\cite{Shi2022}'s coherence-to-displacement conversion and Shapiro's use of sequential detection~\cite{Shapiro2022} to break Nair's performance limit on noise-free target detection~\cite{Nair2011}, we propose a hetero-homodyne receiver for QI, a cascaded POVM that, unlike prior QI receivers, does \emph{not} need a quantum interaction between QI's returned radiation and its stored idler, see Fig.~\ref{Fig_QI}. Our receiver achieves a 3\,dB advantage over the optimum receiver for Tan~\emph{et al}.'s QI, i.e., 9\,dB better than a conventional classical radar.

We will start our development in Sec.~\ref{CIvsQI} with a review of classical versus quantum illumination \emph{without} sequential detection.  Next,  in Sec.~\ref{HetHom}, we will describe our hetero-homodyne receiver and show that it is a cascaded-POVM variant of the PC receiver which achieves the same 3\,dB quantum advantage over classical illumination (CI) when sequential detection is not employed.  Section~\ref{CIQIseq} introduces sequential detection and shows how it affords 6\,dB performance gains to \emph{both} CI and our hetero-homodyne QI when they operate at low single-trial signal-to-noise ratio within the asymptotic regime of interest, i.e., detecting the presence of a weakly-reflecting target embedded in high-brightness background noise.  In Sec.~\ref{simulations} we report simulation results for CI and QI sequential detection that illustrate the distribution of the number of trials needed by these receivers and compare error probabilities obtained from simulations to the earlier asymptotic results.  We wrap up with Sec.~\ref{Discussion}'s summary of what was accomplished plus some additional issues regarding the hetero-homodyne receiver and its use with sequential detection.

\section{Classical versus quantum illumination \label{CIvsQI}}The target-detection scenario of interest is the following.  A positive-frequency, $\sqrt{\mbox{photons/s}}$-units, quantum field operator $\hat{E}_S(t)e^{-i\omega_0t}$ with center frequency $\omega_0$ has its excitation time-limited to $t\in \mathcal{T}_0 \equiv [0,T]$ and (approximately) bandlimited to $B \gg 1/T$\,Hz.  This signal field interrogates a region of space at range $R$ in which there may (hypothesis $H_1$) or may not (hypothesis $H_0$) be a weakly-reflecting target embedded in always-present, high-brightness background radiation.  For $t\in \mathcal{T}_R \equiv [2R/c,2R/c + T]$, the returned radiation collected from that region, after bandlimiting to $B$\,Hz about $\omega_0/2\pi$, is then 
\begin{equation}
\hat{E}_R(t)e^{-i\omega_0t} = \hat{E}^{(0)}_B(t)e^{-i\omega_0t}, 
\end{equation}
under $H_0$, and
\begin{align}
\hat{E}&_R(t)e^{-i\omega_0t} = \nonumber \\[.05in]
&[\sqrt{\kappa}\,e^{i\theta}\hat{E}_S(t-2R/c) + \sqrt{1-\kappa}\,\hat{E}_B^{(1)}(t)]e^{-i\omega_0t}, 
\end{align}
under $H_1$.  Here:  $c$ is light speed; $0 < \kappa \ll 1$ is the roundtrip radar-to-target-to-radar transmissivity; $
\theta$ is the target return's phase delay; and the background noise's baseband field operators, $\hat{E}^{(0)}_B(t)$ and $\hat{E}_B^{(1)}(t)$, are in zero-mean Gaussian states that are completely characterized by their fluorescence spectra~\cite{footnote0}, 
\begin{equation}
S^{(k)}_{BB}(\omega) = \int\!{\rm d}\tau\,\langle \hat{E}_B^{(k)\dagger}(t+\tau)\hat{E}_B^{(k)}(t)\rangle e^{i\omega\tau},
\end{equation}
for $k=0,1$, with~\cite{footnote1} 
\begin{equation}
S_{BB}^{(0)}(\omega) = \left\{\begin{array}{ll} N_B \gg 1, & \mbox{for $|\omega|/2\pi \le B/2$},\\[.05in]
0, & \mbox{elsewhere,}\end{array}\right.
\end{equation}
and 
\begin{equation}
S_{BB}^{(1)}(\omega) = \left\{\begin{array}{ll} N_B/(1-\kappa), & \mbox{for $|\omega|/2\pi \le B/2$},\\[.05in]
0, & \mbox{elsewhere.}\end{array}\right.
\end{equation}    

For simplicity, we shall assume equally-likely hypotheses, set the phase delay $\theta$ to $2\omega_0R/c$~\cite{footnote2}, and take error probability to be our performance metric.  In CI, minimum error-probability operation for a given average transmitted photon number, 
\begin{equation}
N_T \equiv \int_0^T\!{\rm d}t\,\langle \hat{E}_S^\dagger(t)\hat{E}_S(t)\rangle,
\end{equation}
is achieved using coherent-state radiation $|E_S(t)\rangle_S$~\cite{footnote3} with 
\begin{equation}
\int_0^T\!{\rm d}t\, |E_S(t)|^2 = N_T.
\end{equation}
The quantum Chernoff bound~\cite{Calsamiglia2008} on optimum CI's error probability, easily evaluated using results from Pirandola and Lloyd~\cite{Pirandola2008}, is
\begin{equation}
\Pr(e)_{CI} \le e^{-\kappa N_T/4N_B}/2.
\end{equation}  This bound is exponentially tight in increasing signal-to-noise ratio, ${\rm SNR} \equiv \kappa N_T/N_B$.  Indeed, homodyne detection realizes 
\begin{equation}
\Pr(e)^{\rm hom}_{CI} = Q(\sqrt{{\rm SNR}/2})\le e^{-{\rm SNR}/4}/2,
\label{CIhomPr(e)}
\end{equation} where 
\begin{equation}
Q(x) \equiv \int_x^\infty\!{\rm d}y\,e^{-y^2/2}/\sqrt{2\pi},
\end{equation}
and so is the optimum CI receiver when $N_S \ll 1$, $\kappa \ll 1$, and $N_B\gg 1$, which is our asymptotic regime of interest.

Tan~\emph{et al}.'s QI carves duration $T$, bandwidth $B\gg 1/T$ signal and idler pulses from an SPDC's output.  In annihilation-operator mode expansions, their baseband field operators are 
\begin{equation}
\hat{E}_S(t) = \sum_{m=-\infty}^\infty\hat{a}_{S_m}\frac{e^{-i2\pi mt/T}}{\sqrt{T}},
\end{equation}
and 
\begin{equation}
\hat{E}_I(t) = \sum_{m=-\infty}^\infty\hat{a}_{I_m}\frac{e^{i2\pi mt/T}}{\sqrt{T}},
\end{equation} for $t\in \mathcal{T}_0$, and their excited mode pairs, $\{(\hat{a}_{S_m},\hat{a}_{I_m}): -(M-1)/2\le m\le (M-1)/2\}$ with $M\gg 1$ an odd integer, are in iid TMSV states with average photon number $N_S \ll 1$ in each mode, and $MN_S = N_T$.  The signal interrogates the region of interest, as in CI, while the idler is retained for a joint measurement with the returned radiation.  Because this QI setup still involves Gaussian states, the optimum QI receiver's Chernoff bound is easily found to be 
\begin{equation}
\Pr(e)_{QI} \le e^{-\kappa MN_S/N_B}/2,
\end{equation} 
whose error-probability exponent is 6\,dB higher than that of optimum CI.  For the PA and PC receivers, Guha and Erkmen~\cite{Guha2009} show that 
\begin{equation}
\Pr(e)^{\rm PA}_{QI} \approx \Pr(e)^{\rm PC}_{QI} \le e^{-\kappa MN_S/2N_B}/2
\end{equation}
are the relevant Chernoff bounds.  A central limit theorem argument, justified by $M \gg 1$, gives 
\begin{equation}
\Pr(e)^{\rm PA}_{QI} \approx \Pr(e)^{\rm PC}_{QI} \approx Q(\sqrt{\kappa MN_S/N_B}),
\end{equation}
i.e., these receivers achieve a 3\,dB quantum advantage over optimum CI.  

\begin{figure}[t!]
\centering
{\includegraphics[width=0.7\columnwidth]{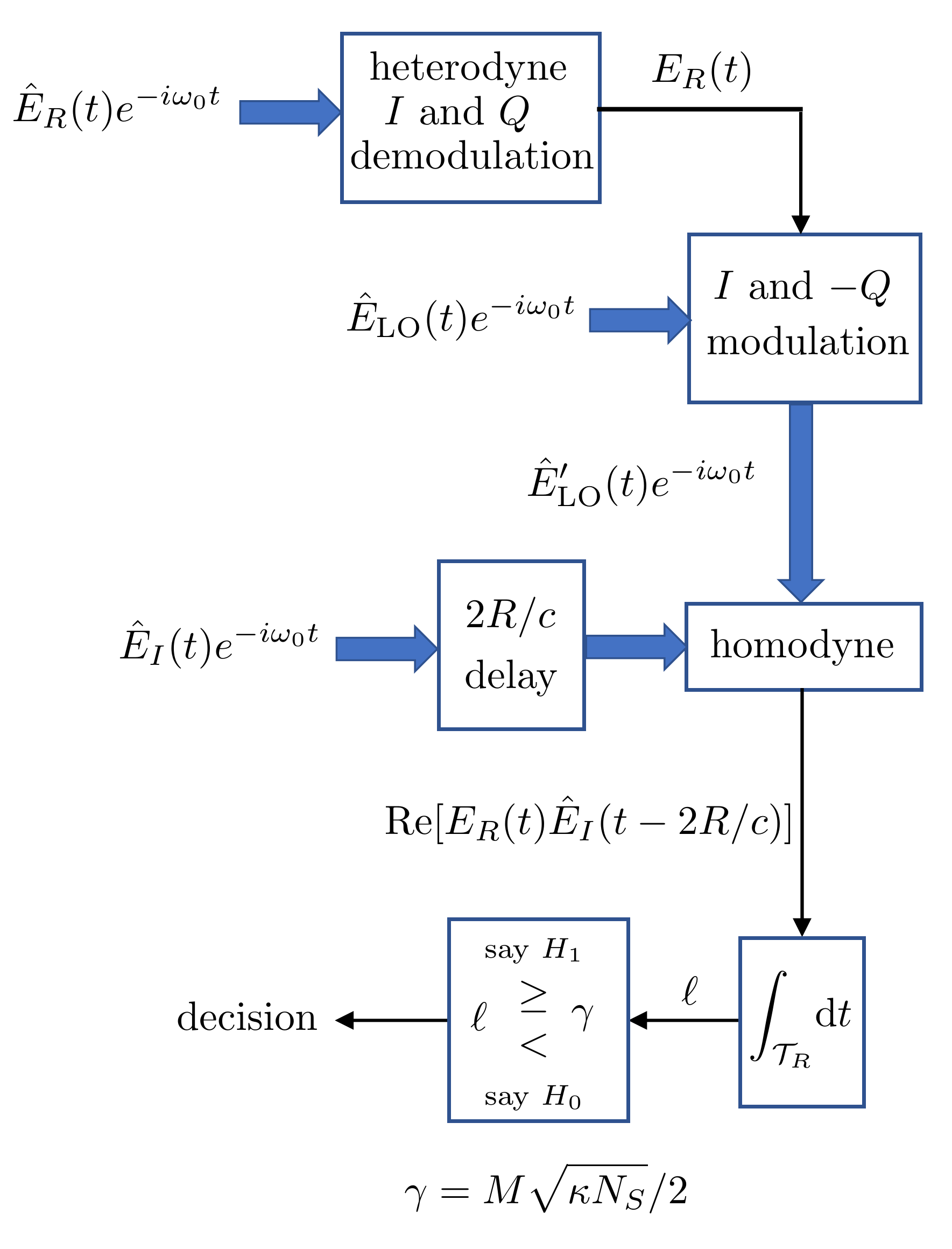}}
\caption{\label{HetHomFig} Schematic of the hetero-homodyne receiver.  Thick arrows represent quantum microwave fields.  Thin arrows represent baseband signals that are conditioned on the outcome of the heterodyne measurement.}
\end{figure}

\section{The hetero-homodyne receiver \label{HetHom}}Figure~\ref{HetHomFig} shows a schematic of the hetero-homodyne receiver.  The $T$-s-duration returned radiation---shown by its positive frequency field operator---undergoes ideal (unit quantum efficiency, bandwidth $B$) heterodyne detection to an intermediate frequency $\omega_{\rm IF}$.  In-phase ($I$) and quadrature ($Q$) demodulation results in the complex-valued measurement outcome $E_R(t)$ whose classical statistics are those of the positive operator-valued measurement (POVM) $\hat{E}_R(t)$~\cite{Shapiro2009}.  That measurement outcome then does $I$ and $-Q$ modulation~\cite{footnote4} of a microwave local oscillator (LO) field, $\hat{E}_{\rm LO}(t)e^{-i\omega_0t}$, with $\hat{E}_{\rm LO}(t)$ in the coherent state $|E_{\rm LO}(t)\rangle$ with mean field
\begin{equation}
\langle \hat{E}_{\rm LO}(t)\rangle = E_{\rm LO}(t) = 
 \sqrt{P_{\rm LO}/T}, \mbox{ for $t\in\mathcal{T}_R$}.
\end{equation}  The resulting LO field, $\hat{E}'_{\rm LO}(t)e^{i\omega_0t}$, is used for ideal (unit quantum efficiency, bilateral bandwidth $B$) homodyne detection of the idler field, which has been held in a quantum memory for $2R/c$, the range delay to the region of interest.  Conditioned on the heterodyne detector's output, $\hat{E}'_{\rm LO}(t)$ will be in a coherent state with   conditional mean 
\begin{equation}
\mathbb{E}[\,\hat{E}'_{\rm LO}(t)\mid E_R(t)\,] \propto E_R^*(t), \mbox{ for $t\in \mathcal{T}_R$}.
\end{equation}
Consequently, the homodyne detector's output, conditioned on the heterodyne detector's output, is---with a convenient normalization---a measurement of the observable ${\rm Re}[E_R(t)\hat{E}_I(t-2R/c)]$ for $t\in \mathcal{T}_R$.  The sufficient statistic $\ell$ for the minimum error-probability processor based on knowledge of the heterodyne detector's output and the observable measurement is the integral of the homodyne output over $\mathcal{T}_R$, and the minimum error-probability decision is a simple threshold test, as shown in Fig.~\ref{HetHomFig}.  The derivation of this optimum decision strategy is given in Appendix~\ref{AppA}; what follows is a simplified performance analysis that assumes $N_S \ll 1$, $\kappa \ll 1$, and $N_B \gg 1$, and uses $M \gg 1$ to justify a central limit theorem approximation.  

To understand the hetero-homodyne receiver, it helps to use the mode expansions given earlier.  Heterodyne detection of the $m$th excited return mode, $\hat{a}_{R_m}e^{-i2\pi m(t-2R/c)/T}\sqrt{T}$ for $t\in \mathcal{T}_R$ and $-(M-1)/2\le m \le (M-1)/2$, followed by $I$ and $Q$ demodulation, yields outcomes equal to the real and imaginary parts of the $\hat{a}_{R_m}$ POVM.  Defining $a_{R_m}$ to be the complex number assembled from those outcomes, we get the baseband waveform 
\begin{equation}
E_R(t-2R/c) = \sum_{m=-(M-1)/2}^{(M-1)/2} a_{R_m}\frac{e^{-i2\pi m(t-2R/c)/T}}{\sqrt{T}}
\end{equation}
as the classical signal applied to the $I$ and $-Q$ modulator.  That modulator transforms the coherent-state local oscillator's mean field $E_{\rm LO}(t)$ to 
\begin{equation}
E'_{\rm LO}(t) \propto \sum_{m=-(M-1)/2}^{(M-1)/2}a^*_{R_m}\frac{e^{-i2\pi m(t-2R/c)/T}}{\sqrt{T}},
\end{equation}
for $t\in \mathcal{T}_R$, \emph{conditioned} on the heterodyne detectors' output.  Moreover, under that conditioning, $\hat{E}'_{\rm LO}(t)$ remains a coherent-state field.  Thus, under that conditioning, integrating the homodyne detector's output over $\mathcal{T}_R$ is (with a convenient normalization) a measurement of 
\begin{align}
\hat{\ell} &\equiv \int_{\mathcal{T}_R}\!{\rm d}t\,{\rm Re}[E_R(t)\hat{E}_I(t-2R/c)] \\[.05in]
& = \sum_{m=-(M-1)/2}^{(M-1)/2}{\rm Re}(a_{R_m}\hat{a}_{I_m}).
\end{align}
The essence of the hetero-homodyne receiver is revealed by this result: the sufficient statistic measures the phase-sensitive cross correlation between the returned radiation and the stored idler, viz., QI's signature of target presence.  At this point, QI performance analysis for the hetero-homodyne reception is easy, as we now show. 

Let $\{\ell_m\}$ be the classical outcomes of the $\{{\rm Re}(a_{R_m}\hat{a}_{I_m})\}$ measurements.  Then, given the true hypothesis and the heterodyne detector's outputs, the fluctuating parts of the $\{\ell_m\}$,
\begin{equation}
\Delta \ell_m \equiv \ell_m - \mathbb{E}(\ell_m \mid H_k, a_{R_m}),
\end{equation}
are iid zero-mean Gaussian random variables.  It follows that the $\{\ell_m\}$ are statistically independent Gaussian random variables given the true hypothesis and the $\{a_{R_m}\}$. (See Appendix~\ref{AppA} for details of the projected idler states created under $H_0$ and $H_1$ by heterodyne detection of the returned radiation.)  Now, because the $\{a_{R_m}\}$ are iid under $H_0$ and $H_1$, we have that the $\{\ell_m\}$ are iid given the true hypothesis. So, because $M \gg 1$, the central limit theorem implies that $\ell$ is Gaussian distributed given the true hypothesis.  

For $N_S \ll1$, $\kappa \ll 1$, and $N_B \gg 1$, the $\hat{\ell}$ measurement's outcome has conditional means
\begin{equation}
\mathbb{E}[\,\ell \mid H_0\,] = 0,
\end{equation}
and 
\begin{equation}
\mathbb{E}[\,\ell \mid H_1\,]  \approx M\sqrt{\kappa N_S},
\end{equation}
and conditional variances 
\begin{equation}
{\rm Var}[\,\ell\mid H_0\,] \approx {\rm Var}[\,\ell \mid H_1\,]  \approx MN_B/4;
\end{equation}
see Appendix~\ref{AppA} for the exact results.  It follows that the minimum error-probability decision rule is to decide $H_1$ when 
\begin{equation}
\ell \ge \gamma \equiv M\sqrt{\kappa N_S}/2
\end{equation}
and decide $H_0$ otherwise.  This rule's error probability is 
\begin{equation}
\Pr(e) \approx Q(\sqrt{M\kappa N_S/N_B}) \le e^{-M\kappa N_S/2N_B}/2,
\label{QILRTPr(e)}
\end{equation} 
whose error-probability exponent is 3\,dB better than that of the best CI radar with $N_T = MN_S$.  

It is no accident that the hetero-homodyne receiver achieves the same performance as Guha and Erkmen's PC receiver.  As seen in Fig~\ref{HetHomFig}, heterodyne detection followed by $I$ and $-Q$ modulation realizes the phase conjugation operation needed to convert the phase-sensitive cross correlation between the returned radiation and the idler into a quantity  measurable via homodyne detection.  The PC receiver uses a parametric amplifier to accomplish phase conjugation that permits the phase-sensitive cross correlation to be observed via conventional second-order interference.  The hetero-homodyne receiver differs from the PC receiver in that the latter measures a POVM on the joint Hilbert space of the returned radiation and the idler, whereas the former measures a POVM on the Hilbert space of the returned radiation and uses the outcome of that measurement to choose the observable that will be measured on the idler.  As a result, the hetero-homodyne receiver suffers less added quantum noise from its conjugation operation than does the PC receiver from its conjugation operation, but both amounts are inconsequential in the asymptotic regime of interest wherein $N_S \ll 1$, $\kappa \ll 1$, and $N_B \gg 1$; see Appendix~\ref{AppB} for the details. Note that QI's hetero-homodyne, PC, and PA receivers all require knowledge of the target return's phase delay $\theta$, which is an issue we shall return to later, as well as a quantum memory to store the idler for the roundtrip range delay $2R/c$.  

It might seem that the hetero-homodyne receiver's requiring a quantum memory for idler storage could be circumvented by: (1) heterodyning the idler, (2) delaying (by $2R/c)$ the \emph{classical} outcome of that measurement, (3) using that delayed signal to $I$ and $-Q$ modulate the LO that homodynes the returned radiation, and (4) performs the likelihood-ratio test (LRT) on the homodyne output.  As shown in Appendix~\ref{AppC}, this approach fails to offer any quantum advantage owing to the idler's state being dominated by quantum noise, whereas the returned radiation's state is dominated by classical noise.  

\section{CI and QI with sequential detection \label{CIQIseq}}Wald~\cite{Wald1945} originated the sequential probability-ratio test (SPRT) as an alternative to the standard (LRT) used for fixed-length data.  Consider observation of a random vector ${\bf x}$ that has conditional probability density functions (pdfs) $p_{{\bf x}\mid H_k}({\bf X}\mid H_k)$ for $k = 0,1$.  The LRT
\begin{equation}
\Lambda_{\bf x}({\bf X}) \equiv \frac{p_{{\bf x}\mid H_1}({\bf X}\mid H_1)}{p_{{\bf x}\mid H_0}({\bf X}\mid H_0)} 
\begin{array}{c}\mbox{\scriptsize say $H_1$}\\ \ge \\ < \\ \mbox{\scriptsize say $H_0$}\end{array} \eta,
\end{equation}
with $\eta = \Pr(H_0)/ \Pr(H_1)$ minimizes the error probability.  Alternatively, $\eta$ can be chosen to ensure that the miss probability, $P_M \equiv \Pr(\mbox{decide $H_0$}\mid \mbox{$H_1$ true})$, is minimized subject to the false-alarm probability constraint $P_F \equiv \Pr(\mbox{decide $H_1$}\mid \mbox{$H_0$ true}) \le \alpha$.  

In a non-adaptive SPRT, a sequence of iid random vectors, $\{{\bf x}_n:  n = 1,2, \ldots\}$, are available, and desired maximum values $\alpha$ and $\beta$ have been set in advance for $P_F$ and $P_M$, respectively.  From $\alpha$, $\beta$, thresholds $A = (1-\beta)/\alpha$ and $B=(1-\alpha)/\beta$ can be set~\cite{Wald1945} such that the following protocol guarantees $P_F \le \alpha$ and $P_M \le \beta$:  
\begin{enumerate}
\item Define ${\bf z}_n = \{{\bf x}_1,{\bf x}_2,\ldots,{\bf x}_n\}$. 

\item For $n=1,2,\ldots,$ until the protocol terminates, compute $\Lambda_{{\bf z}_n}({\bf Z}_n)$.  

\item If $\Lambda_{{\bf z}_n}({\bf Z}_n) \ge A$, declare $H_1$ and terminate.  

\item If $\Lambda_{{\bf z}_n}({\bf Z}_n) \le B$, declare $H_0$ and terminate. 

\item If $B < \Lambda_{{\bf z}_n}({\bf Z}_n) < A$, increment $n$ and continue.  
\end{enumerate}
Furthermore, at low single-trial SNR Ref.~\cite{Wald1945} shows that $P_F \approx \alpha$ and $P_M\approx \beta$.

To exhibit the benefits offered by sequential detection, we shall assume $N_S \ll 1$, $\kappa \ll 1$ and $N_B \gg 1$, and apply the SPRT protocol to CI using homodyne detection and to QI using hetero-homodyne detection.  For each transmission, the CI transmitter will employ a coherent state $|\sqrt{M_SN_S}\rangle$. The QI transmitter, on the other hand, will employ  $M_S$ iid TMSV mode pairs with average photon number $N_S$ in each signal and idler.  For both systems we assume $M_S \gg 1$ and $\kappa M_SN_S/N_B \ll 1$~\cite{footnote5} and evaluate the average number of transmissions---hence the average transmitted photon number---under the assumptions of equally-likely hypotheses and $\alpha=\beta$, i.e., equal false-alarm and miss probabilities.  Within this operating regime, the results we seek follow readily from Ref.~\cite{Wald1945}.   

Consider first CI's SPRT.  On each transmission the homodyne receiver's sufficient statistic, $\ell_n^{\rm hom}$, is a conditionally-Gaussian random variable with conditional means 
\begin{equation}
\mathbb{E}[\,\ell_n^{CI} \mid H_0\,] = 0
\end{equation}
and 
\begin{equation}
\mathbb{E}[\,\ell_n^{CI} \mid H_1\,] = M_S\sqrt{\kappa N_S},
\end{equation}
and conditional variances 
\begin{equation}
{\rm Var}[\,\ell_n^{CI}\mid H_0\,] = {\rm Var}[\,\ell_n^{CI}\mid H_1\,] = M_SN_B/2.
\end{equation}  
Let $K_{CI}$ be the number of trials at which the SPRT protocol terminates.  From the result in Sec.~5.4.5 of Ref.~\cite{Wald1945}, we have that
\begin{align}
\mathbb{E}[\,K_{CI}&\mid H_0\,] \approx \mathbb{E}[\,K_{CI}\mid H_1\,] \\[.05in] &\approx  N_B\ln[(1-\alpha)/\alpha](1-2\alpha)/\kappa M_SN_S,
\end{align} 
under our assumption of low single-trial SNR.  It then follows that the average number of transmissions  is 
\begin{equation}
\langle K_{CI}\rangle  \approx N_B\ln[(1-\alpha)/\alpha](1-2\alpha)/\kappa M_SN_S,
\label{Kci}
\end{equation}
and the average number of transmitted photons is $N^{\rm SPRT}_T = \langle K_{CI}\rangle M_SN_S$.

With $\alpha = \beta$, our CI SPRT is guaranteed to have $\Pr(e)\le \alpha$, and this bound is tight at low single-trial SNR~\cite{Wald1945}.  So, for operation at $\Pr(e) \ll 1$, we can use 
\begin{equation}
\Pr(e) \sim e^{-\kappa N^{\rm LRT}_T/4N_B}
\end{equation}
for LRT-based CI target detection, i.e., without sequential detection a low error probability requires 
\begin{equation}
N^{\rm LRT}_T \sim -4N_B\ln[\Pr(e)]/\kappa
\end{equation}
photons on average.  SPRT-based CI target detection, however, achieves the \emph{same} low error probability with  
\begin{equation}
N^{\rm SPRT}_T  = N^{\rm LRT}_T/4,
\end{equation}
viz., a 6\,dB sequential-detection advantage in error-probability exponent.  Putting aside operational considerations regarding sequential detection's practicality for radar target detection---see Ref.~\cite{Shapiro2022} for some discussion of these considerations---this CI sequential-detection advantage matches the quantum advantage of the yet-to-be implemented optimum quantum receiver for QI without sequential detection.  Thus, if sequential detection is suitable for radar target detection, there should be little interest in further pursuit of Tan~\emph{et al}.\@ QI without sequential detection.

Now consider SPRT-based QI target detection using the hetero-homodyne receiver.  Similar to what we have done earlier for QI without sequential detection, we can use $N_S \ll 1$, $\kappa \ll 1$, $N_B \gg 1$, and $M_S \gg 1$, to justify approximating the receiver's single-trial sufficient statistics, $\{\ell_n^{QI}\}$, as iid conditionally-Gaussian random variables with conditional means 
\begin{equation}
\mathbb{E}[\,\ell_n^{QI} \mid H_0\,] =0
\end{equation}
and
\begin{equation}
\mathbb{E}[\,\ell_n^{QI} \mid H_1\,] \approx M_S\sqrt{\kappa N_S},
\end{equation}
and conditional variances
\begin{equation}
{\rm Var}[\,\ell_n^{QI}\mid H_0\,] \approx {\rm Var}[\,\ell_n^{QI} \mid H_1\,] \approx M_SN_B/4.
\end{equation}  
Then, because we again have low single-trial SNR, we can parallel what we did for SPRT-based CI and find that 
\begin{equation}
N_T^{\rm SPRT} = N_T^{\rm LRT}/4,
\end{equation}
where the left-hand side is the average number of transmitted photons needed to achieve a particular error probability using hetero-homodyne QI reception with an SPRT, and the right-hand side is the average number of transmitted photons needed to realize that same error probability using hetero-homodyne reception with a single-transmission LRT.  Here too there is a 6\,dB sequential-detection advantage in error-probability exponent for SPRT QI versus LRT QI.  Also, as was the case for LRT QI versus LRT CI, we find that SPRT QI has a 3\,dB quantum advantage over SPRT CI.  Note that compared to conventional (non-sequential) CI, hetero-homodyne QI with sequential detection provides a 9\,dB quantum advantage.  

\section{Simulation Results \label{simulations}}
Here we present sequential-detection simulation results for CI and QI illustrating the probability distributions for their number of required trials and comparing their error probabilities with those of their non-sequential counterparts.

\begin{figure}[t]
\centering
{\includegraphics{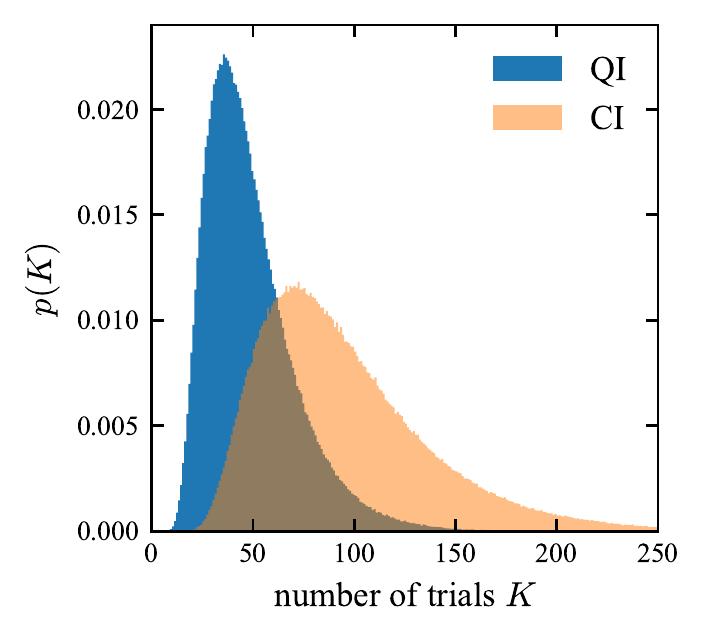}}
\caption{\label{p(K)} Simulated probability distributions for the number of SPRT trials used by CI and QI assuming  $N_S = 0.01$, $\kappa = 0.01$, $N_B = 100$, $M_S = 10^5$, and $\alpha = \beta = 10^{-4}$.  Each distribution was generated from $10^6$ simulated experiments for each hypothesis.}
\end{figure}
Figure~\ref{p(K)} shows the probability distributions for the number of SPRT trials used by CI and QI assuming $N_S = 0.01$, $\kappa = 0.01$, $N_B = 100$, $M_S = 10^5$, and $\alpha = \beta = 10^{-4}$, with each distribution being generated from $10^6$ simulated experiments for each hypothesis.  These parameter values imply operation at the edge of low single-trial SNR, viz., $M\kappa N_S/N_B = 0.1$, thus appreciable deviations from asymptotic behavior might occur.  So, because 
\begin{equation}
{\Pr}_{\rm SPRT}^{CI}(e) \sim e^{-\langle K_{CI}\rangle M_S\kappa N_S/N_B}
\label{ChernoffCI}
\end{equation}
in this regime, we expect $\langle K_{CI}\rangle \sim 92$ for ${\Pr}_{\rm SPRT}^{CI}(e) \approx \alpha = \beta = 10^{-4}$, which is in good agreement with the simulation result $\langle K_{CI}\rangle = 95$.  The simulation gave $\log_{10}[{\Pr}_{\rm SPRT}^{CI}(e)] = -4.13 \pm 0.073$ with 95\% confidence, implying $6.21 \times 10^{-5} \le {\Pr}_{\rm SPRT}^{CI}(e) \le 8.70 \times 10^{-5}$, in good agreement with theory given that this example is at the edge of the low single-trial SNR regime.

Turning now to the QI case,
\begin{equation}
{\Pr}_{\rm SPRT}^{QI}(e) \sim e^{-2\langle K_{QI}\rangle M_S\kappa N_S/N_B}
\label{ChernoffQI}
\end{equation}
leads us to expect $\langle K_{QI}\rangle \sim 46$ for ${\Pr}_{\rm SPRT}^{QI}(e) \approx \alpha = \beta = 10^{-4}$, which is in good agreement with our simulation result $\langle K_{QI}\rangle = 49$.  The simulation gave $\log_{10}[{\Pr}_{\rm SPRT}^{QI}(e)] = -4.17 \pm 0.077$ with 95\% confidence, implying $5.66 \times 10^{-5} \le {\Pr}_{\rm SPRT}^{QI}(e) \le 7.93 \times 10^{-5}$.  Again we see good agreement with theory given operation is at the edge of the low single-trial SNR regime.

\begin{figure}[t]
\centering
{\includegraphics{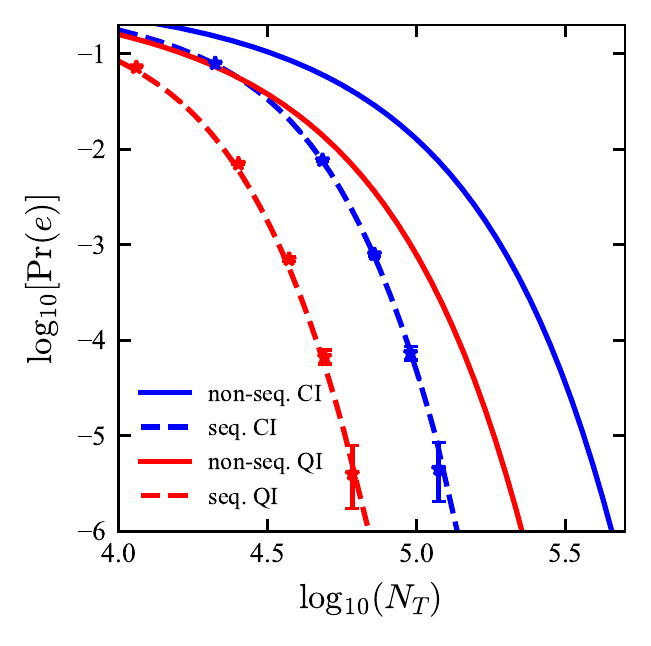}}
\caption{\label{Pr(e)} CI and QI error probabilities versus average number of transmitted signal photons, $N_T$, assuming $N_S = 0.01$, $\kappa = 0.01$, and $N_B = 100$. The solid curves are theory results from Eqs.~(\ref{CIhomPr(e)}) and (\ref{QILRTPr(e)}) for non-sequential CI with homodyne detection and non-sequential QI with hetero-homodyne detection, respectively.  The points are simulated error probabilities for sequential CI with homodyne detection and sequential QI with hetero-homodyne detection with $M_S = 10^5$ for $\alpha = \beta = 10^{-1}, 10^{-2}, 10^{-3}, 10^{-4}$, and $10^{-5}$.  Each point was generated from $10^6$ simulated experiments for each hypothesis. See the text for how the dashed curves were obtained.}
\end{figure}
Figure~\ref{Pr(e)} compares the error probabilities of non-sequential and sequential CI and QI versus their average number of transmitted signal photons, $N_T$.  All assume $N_S = 0.01$, $\kappa = 0.01$, and $N_B = 100$ with CI using homodyne detection and QI using hetero-homodyne detection.  The solid curves are theoretical results from Eqs.~(\ref{CIhomPr(e)}) and (\ref{QILRTPr(e)}) for CI and QI, respectively.  The points are sequential results from $10^6$ experiments for each hypothesis using $M_S = 10^5$ with $\alpha=\beta = 10^{-1},10^{-2},10^{-3},10^{-4}$, and $10^{-5}$.  The dashed curves were obtained as follows.  For sequential CI we used $\Pr_{\rm SPRT}^{CI}(e) \approx \alpha$ in Eq.~(\ref{Kci}) and solved for $\langle K_{CI}\rangle$ as a function of $\Pr_{\rm SPRT}^{CI}(e)$.  Then we plotted $\Pr_{\rm SPRT}^{CI}(e)$ versus 
\begin{equation}
N_T \approx N_B\ln\!\left[\frac{[1-{\Pr}_{\rm SPRT}^{CI}(e)]}{{\Pr}_{\rm SPRT}^{CI}(e)}\right]\frac{[1-2{\Pr}_{\rm SPRT}^{CI}(e)]}{\kappa}.
\end{equation} 
Sequential QI's dashed curve was obtained, using its 3\,dB energy advantage over sequential CI, from
\begin{equation}
N_T \approx N_B\ln\!\left[\frac{[1-{\Pr}_{\rm SPRT}^{QI}(e)]}{{\Pr}_{\rm SPRT}^{QI}(e)}\right]\frac{[1-2{\Pr}_{\rm SPRT}^{QI}(e)]}{2\kappa}.
\end{equation} 

Examination of Fig.~\ref{Pr(e)} shows excellent agreement between theory and simulation for sequential CI and QI, with the large 95\% confidence interval for $\alpha = \beta = 10^ {-5}$ being due to our using only $10^6$ experiments for each hypothesis in the simulations.  Figure~\ref{Pr(e)} also confirms the 6\,dB error-exponent advantage that sequential CI and sequential QI enjoy over their non-sequential counterparts.

\section{Discussion \label{Discussion}}Our paper has made two significant advances for microwave QI:  (1) it proposed the hetero-homodyne receiver, whose non-sequential performance matches that of the Guha and Erkmen's PA and PC receivers in QI's usual $N_S \ll 1$, $\kappa \ll 1$, $N_B \gg 1$, $M \gg 1$ operating regime, but exceeds their performance outside this regime; and (2) it showed that sequential detection, at low single-trial SNR in QI's usual operating regime, provides a 6\,dB increase in error-probability exponent for both CI homodyne detection and QI hetero-homodyne detection
as compared to their non-sequential counterparts.  In addition, it showed that cascaded POVMs---of which the hetero-homodyne receiver is an example---should be considered for quantum radar.

Advance (2) demonstrates that Tan~\emph{et al}.'s TMSV QI does \emph{not} saturate what can be gained from using entanglement for target detection in a lossy, noisy environment despite the optimality proofs from Refs.~\cite{DePalma2018,Nair2020,DiCandia2021,Bradshaw2021, Sanz2017, Gong2022,Jonsson2022}. That demonstration begs the question of why it does not contradict those proofs.  The answer is simple.  Those proofs assume pure-state, signal-idler pulses with fixed time duration, whereas sequential detection employs a random number of fixed-duration pulses making its overall time duration random. 

 Three other questions that immediately arise are: (1) whether the $M_S$ mode-pair TMSV states are optimum for hetero-homodyne reception using our sequential protocol; (2) whether known optimum QI receivers for single-pulse operation can increase their 6\,dB quantum advantage by employing them in the sequential protocol; and (3) for either single-pulse receiver, whether our form of the sequential protocol can be improved. With regard to the first question, we suspect that TMSV states are optimal for the hetero-homodyne reception’s non-adaptive SPRT, but a different answer might emerge from recent work on quantum sequential detection~\cite{MartinezVargas2021}. The answer to the second question is a definite yes. Replacing the hetero-homodyne receiver with either the feed-forward sum-frequency generation receiver~\cite{Zhuang2017} or the coherence-to-displacement receiver~\cite{Shi2022} improves the SNR on each pulse by 3\,dB in comparison with hetero-homodyne reception. So, operating at low single-pulse SNR, these optimum receivers will accrue an additional 6\,dB quantum advantage over their non-sequential counterparts, as can easily be proved by paralleling the calculation we did for hetero-homodyne reception’s SPRT.  Of course, this performance gain comes at the expense of vastly more complicated single-pulse receivers. For the third question we note that sequential use of the hetero-homodyne receiver or the optimum QI receivers might increase their quantum advantages by letting their $k^{th}$ transmitted pulse employ a brightness, $N_S(k)$, that depends on the likelihood accumulated from the prior measurements in a manner that minimizes the average number of pulses needed to reach convergence. We leave exploration of this possibility for future work.

At this point it behooves us to address some additional issues regarding the hetero-homodyne receiver and its use with sequential detection.  On the negative side are its continuing need for knowledge of the target's phase delay and its requiring a quantum memory for idler storage.  Conventional (non-sequential) CI with homodyne detection has no idler to store, but it has the same  need for phase-delay information. That said, CI target detection employing heterodyne detection, plus matched filtering and envelope detection at the intermediate frequency, obviates the need for phase-delay information and suffers only $\sim$3\,dB loss in error-probability exponent~\cite{VanTrees2001}.  Non-sequential QI with a dual hetero-homodyne receiver that homodynes both quadratures and incoherently combines their outputs offers insensitivity to the target's phase delay, but it suffers a 3\,dB SNR loss from its 50--50 splitting of the stored idler and therefore fails to provide any quantum advantage.

On the positive side for sequential hetero-homodyne QI is its less demanding bandwidth requirement.  In particular, a low QI error probability, $\Pr(e) \ll 1$, requires 
\begin{equation}
B \sim -2N_B\ln[\Pr(e)]/\kappa N_ST
\end{equation} 
for non-sequential operation and 
\begin{equation}
B_S \sim -N_B\ln[\Pr(e)]/2\kappa N_ST\langle K_{QI}\rangle \ll B
\end{equation}
for sequential operation at the same pulse duration.  Because we are operating with $N_S \ll 1$, $\kappa \ll 1$ and $N _B \gg 1$, both of these bandwidth requirements will be demanding in the microwave region~\cite{Jonnson2021}, but that for sequential operation will be substantially less so.  In fairness, we must note that error probability at fixed $\kappa N_S/N_B$ is determined by time-bandwidth product, and the sequential approach's relaxed bandwidth requirement comes from its using a total time duration in excess of $\langle K_{QI}\rangle T$\, as compared to non-sequential operation's $T$\,s duration.  So non-sequential operation that does coherent processing of $M/M_S$ bandwidth-$B_S$ pulses would use a total time duration equal to the sequential system's average time duration.  In that case, although both systems use the same effective time-bandwidth product, the multi-pulse non-sequential system would have 6\,dB lower SNR than its sequential counterpart, as seen in Sec.~\ref{CIQIseq}.

A final point worth noting is that the hetero-homodyne receiver does not circumvent QI's single-bin-interrogation limit, i.e., for best performance its receiver's joint measurement must be made on returned radiation from a single azimuth-elevation-range-Doppler-polarization bin~\cite{Shapiro2020}.  Hence, the total time duration spent probing that bin must be short enough to freeze the target's motion.  Sequential detection's entailing longer interrogation times than non-sequential (single-pulse) operation exacerbates this problem for detecting airborne targets, and especially so when it utilizes far more than its average number of pulses; see\@ Sec.~\ref{simulations}'s distributions for the number of trials needed. However, hetero-homodyne reception is applicable to tasks that are not plagued by the preceding limits on its use in quantum radar. Indeed, we believe its non-adaptive form is immediately relevant to the phase estimation,  entanglement-assisted communication, and general thermal-loss channel pattern hypothesis testing considered by Shi et al.~\cite{Shi2022}.  For example, hetero-homodyne reception’s same non-sequential LRT can be used, with $\gamma = 0$, for entanglement-assisted binary phase-shift keyed communication.

In conclusion, we believe the hetero-homodyne receiver we have proposed both pushes microwave QI target detection closer to fruition and underscores the need for continued research into quantum radar.  

\acknowledgments
We thank Robert Jonsson for his valuable insights on a potential microwave implementation of the hetero-homodyne receiver. M.R. acknowledges support from the UPV/EHU PhD Grant PIF21/289 and from the QUANTEK project from ELKARTEK program (KK-2021/00070). Q.Z. acknowledges support from National Science Foundation CAREER Award CCF-2142882, Office of Naval Research Grant No. N00014-23-1-2296 and Cisco Systems, Inc. J.H.S. acknowledges support from the MITRE Corporation's Quantum Moonshot Program. R.D. acknowledges support from the Academy of Finland, grants no. 353832 and 349199.

\appendix
\section{Likelihood-Ratio Test for Hetero-Homodyne Reception \label{AppA}}
Our first task, in this appendix, is to show that with $E_R(t)$ for $t \in \mathcal{T}_R$ being the classical random process outcome from heterodyne detection of $\hat{E}_R(t)$, and 
\begin{equation}
\hat{\ell} \equiv \int_{\mathcal{T}_R}\!{\rm d}t\,{\rm Re}[E_R(t)\hat{E}_I(t-2R/c)]
\end{equation}
being an observable obtained from homodyne detection of $\hat{E}_I(t-2R/c)$ using a local oscillator whose mean field is proportional to $E_R^*(t)$, then the LRT for hetero-homodyne QI's minimum error-probability decision between equally-likely target absence and presence reduces to threshold test
\begin{equation}
\ell \begin{array}{c} \mbox{\scriptsize say $H_1$} \\
\ge \\ < \\ \mbox{\scriptsize say $H_0$} \end{array} \gamma 
 \equiv M\sqrt{\kappa N_S}/2,
\end{equation}
where $\ell$ is the outcome of the $\hat{\ell}$ measurement and operation is in QI's usual asymptotic regime, $N_S \ll 1$, $\kappa \ll 1$, $N_B \gg 1$, and $M \gg 1$. The second task, for this appendix, is to validate the threshold test's error probability expression given in Sec.~\ref{HetHom}.

Given the true hypothesis, the $M$ mode pairs, $\{(\hat{a}_{R_m},\hat{a}_{I_m})\}$, comprising $\hat{E}_R(t)$ and $\hat{E}_I(t-2R/c)$ are iid. Thus it suffices to start with just a generic mode pair, $(\hat{a}_R,\hat{a}_I)$, to accomplish our tasks.  The pair's joint density operator, $\hat{\rho}^{(k)}_{RI}$, given hypothesis $H_k$ is true, is completely characterized by its normally-ordered form,
\begin{equation}
\rho_{RI}^{(k)}(a_R,a_I) \equiv {}_R\langle a_R|{}_{\,I}\langle a_I|\hat{\rho}^{(k)}_{RI}|a_I\rangle_I|a_R\rangle_R,
\end{equation}
where $|a_R\rangle_R$ and $|a_I\rangle_I$ are coherent states of the $\hat{a}_R$ and $\hat{a}_I$ modes.  The conditional state of the $\hat{a}_I$ mode under $H_k$, given that heterodyne detection of the $\hat{a}_R$ mode has resulted in outcome $a_R$, is completely characterized by its normally-ordered form
\begin{equation}
\rho_{I\mid R}^{(k)}(a_I\mid a_R) = \frac{{}_R\langle a_R|{}_{\,I}\langle a_I|\hat{\rho}^{(k)}_{RI}|a_I\rangle_I|a_R\rangle_R}
{{}_R\langle a_R|\hat{\rho}^{(k)}_R|a_R\rangle_R},
\label{condxstate1}
\end{equation}
where $\hat{\rho}^{(k)}_R$ is the reduced density operator of the $\hat{a}_R$ mode under hypothesis $H_k$.  
  
For Tan~\emph{et al}.'s target detection scenario, both the numerator and denominator in Eq.~\eqref{condxstate1} represent zero-mean Gaussian states.  Under hypothesis $H_0$, the $\hat{a}_R$ and $\hat{a}_I$ modes are in a product of thermal states with average photon numbers $N_B$ and $N_S$, respectively.  Thus $\rho^{(0)}_{I\mid R}(a_I\mid a_R)$ is a thermal state with average photon number $N_S$, which can be approximated by the vacuum state when $N_S \ll 1$.  Under hypothesis $H_1$, however, the $\hat{a}_R$ and $\hat{a}_I$ modes are correlated, as seen from $\rho^{(1)}_{RI}(\alpha_R,\alpha_I)$'s covariance matrix,
\begin{equation}
\Lambda^{(1)}_{RI} = \left[\begin{array}{cc}
\Lambda^{(1)}_R & \Lambda^{(1)}_{RI} \\[.05in]
(\Lambda^{(1)}_{RI})^T & \Lambda_I \end{array}\right],
\end{equation}
where
\begin{align}
\Lambda^{(1)}_R &= \frac{1}{2}\left[\begin{array}{cc}
\kappa N_S + N_B + 1 & 0 \\[.05in]
0 & \kappa N_S + N_B + 1 \end{array}\right] \\[.05in] &\approx \frac{N_B}{2}\,{\bf I}_2,
\end{align}
\begin{align}
\Lambda^{(1)}_{RI} &= \frac{1}{2}\left[\begin{array}{cc}
\sqrt{\kappa N_S(N_S+1)} & 0 \\[.05in]
0 & -\sqrt{\kappa N_S(N_S+1)}  \end{array}\right] \\[.05in]
&\approx \frac{1}{2}\left[\begin{array}{cc}
\sqrt{\kappa N_S} & 0 \\[.05in]
0 & -\sqrt{\kappa N_S}\end{array}\right],
\end{align}
and
\begin{equation}
\Lambda_I = \frac{N_S+1}{2}\,{\bf I}_2 \approx \frac{{\bf I}_2}{2},
\end{equation}
with ${\bf I}_2$ being the $2\times 2$ identity matrix.

Standard results for jointly-Gaussian random variables now gives us that 
$\rho_{I\mid R}^{(1)}(a_I\mid a_R)/\pi$ is the probability density for 
a  2D Gaussian random vector, 
\begin{equation}
{\boldsymbol a}_I = \left[\begin{array}{c}
{\rm Re}(a_I)  \\[.05in]
{\rm Im}(a_I) \end{array}\right],
\end{equation}
with mean value (using the obvious definition for ${\boldsymbol a}_R$)
\begin{align}
\mathbb{E}[{\boldsymbol a}_I\mid H_1, {\boldsymbol a}_R] &= (\Lambda^{(1)}_{RI})^T(\Lambda^{(1)}_{R})^{-1}{\boldsymbol a}_R, \\[.05in]
&= \frac{\sqrt{\kappa N_S(N_S+1)}}{\kappa N_S + N_B +1}\left[\begin{array}{cc}1 & 0 \\[.05in]
0 & -1\end{array}\right]{\boldsymbol a}_R \\[.05in]
&\approx \frac{\sqrt{\kappa N_S}}{N_B}\left[\begin{array}{cc}1 & 0 \\[.05in]
0 & -1\end{array}\right]{\boldsymbol a}_R,
\end{align}
and covariance matrix
\begin{align}
\Lambda^{(1)}_{I\mid R} &= \Lambda^{(1)}_I - (\Lambda^{(1)}_{RI})^T(\Lambda^{(1)}_{R})^{-1}\Lambda^{(1)}_{RI} \\[.05in]
& = \frac{\displaystyle (N_B+1)(N_S+1)}{\displaystyle 2(\kappa N_S + N_B + 1)}\,{\bf I}_2 \approx \frac{{\bf I}_2}{2}.
\end{align}
Stated succinctly, these results imply that in QI's asymptotic regime $\hat{\rho}^{(1)}_{I\mid R}$ is the coherent state $|(\sqrt{\kappa N_S}/N_B)a_R^*\rangle_I$, where $a_R$ is the outcome of heterodyne detecting the $\hat{a}_R$ mode, while $\hat{\rho}^{(0)}_{I\mid R}$ is the vacuum state.

Given $a_R$, let $\hat{a}'_{\rm LO}$ be the LO mode that beats against the $\hat{a}_I$ mode during homodyne detection.  Under that conditioning the LO mode is in a coherent state $|a'_{\rm LO}\rangle$ with $a'_{\rm LO}\propto a_R^*$.  Thus, with the appropriate normalization, the homodyne output $a'_{I_r}$ can be taken to be ${\rm Re}(a_R\hat{a}_I)$, where the $\hat{a}_I$ mode is in the vacuum state if the target is absent and the coherent state $|(\sqrt{\kappa N_S}/N_B)a_R^*\rangle_I$ if the target is present.  Armed with this information we can now evaluate the LRT for hetero-homodyne QI.  We start from
\begin{equation}
\frac{p_{a_R\mid H_1}(\alpha_R\mid H_1)p_{a'_{I_r}\mid H_1,a_R}(\alpha'_{I_r}\mid H_1,a_R)}{p_{a_R\mid H_0}(\alpha_R\mid H_0)p_{a'_{I_r}\mid H_0,a_R}(\alpha'_{I_r}\mid H_0,a_R)} \begin{array}{c}
\mbox{\scriptsize say $H_1$}\\ \ge \\ < \\ \mbox{\scriptsize $H_0$} \end{array} 1,
\end{equation}
where the threshold value assumes equally-likely target absence and presence.
Substituting in the probability densities, taking logarithms of both sides, and rearranging terms reduces the LRT to 
\begin{equation}
\alpha'_{I_r} \begin{array}{c}
\mbox{\scriptsize say $H_1$}\\ \ge \\ < \\ \mbox{\scriptsize $H_0$} \end{array} 
\frac{\sqrt{\kappa  N_S}}{2N_B}\,|a_R|^2.
\end{equation}
for a decision based on one mode pair.  Summing over all $M$ mode pairs, this threshold test becomes the result we are seeking, viz., 
\begin{equation}
\ell \begin{array}{c}
\mbox{\scriptsize say $H_1$}\\ \ge \\ < \\ \mbox{\scriptsize $H_0$} \end{array} \frac{\sqrt{\kappa N_S}}{2N_B}\int_{\mathcal{T}_R}\!{\rm d}t\,|E_R(t)|^2 \approx M\sqrt{\kappa N_S}/2,
\end{equation}
where the approximation is due to $M \gg 1$ and the law of large numbers.
So, to validate the error probability expression from Sec.~\ref{HetHom}, it only remains for us to verify the asymptotic-regime conditional means and variances of $\ell$ given the true hypothesis.  Here, for completeness, we will postpone assuming $N_S\ll 1$, $\kappa \ll 1$, and $N_B \gg 1$ so that general results are available for use in Appendix~\ref{AppB}.

Using mode expansions for $E_R(t)$ and $\hat{E}_I(t-2R/c)$, we have that
\begin{equation}
\hat{\ell} = \sum_{m=-(M-1)/2}^{(M-1)/2}{\rm Re}(a_{R_m}\hat{a}_{I_m}),
\end{equation}
from which our earlier results immediately give
\begin{equation}
\mathbb{E}(\ell\mid H_0) = 0,
\end{equation}
and
\begin{align}
\mathbb{E}(\ell\mid H_1) &= \sum_{m=-(M-1)/2}^{(M-1)/2}{\rm Re}\left[\frac{\sqrt{\kappa N_S(N_S + 1)}}{\kappa N_S+N_B +1}\langle |a_{R_m}|^2\rangle\right]\\[.05in]
&= M\sqrt{\kappa N_S(N_S+1)} \approx M\sqrt{\kappa N_S},
\end{align} 
which reproduce the conditional means given in Sec.~\ref{HetHom}.

To obtain the $\ell$'s conditional variances, given the true hypothesis, we use iterated expectation, i.e.,
\begin{align}
{\rm Var}(\ell \mid H_k) &= \mathbb{E}\{{\rm Var}[\ell \mid H_k, E_R(t)] \mid H_k\}
\nonumber \\[.05in]
& + {\rm Var}\{\mathbb{E}[\ell \mid H_k, E_R(t)]\mid H_k\}.
\end{align}
Under $H_0$ this result becomes
\begin{equation}
{\rm Var}(\ell \mid H_0) = \sum_{m=-(M-1)/2}^{(M-1)/2}\frac{\langle |a_{R_m}|^2\rangle}{4} = MN_B/4,
\end{equation}
while under $H_1$ it becomes
\begin{eqnarray}
\lefteqn{{\rm Var}(\ell \mid H_1) = \sum_{m=-(M-1)/2}^{(M-1)/2}\frac{\langle |a_{R_m}|^2\rangle}{4} }\nonumber \\[.05in] &\hspace{.2in}&+\,\,\frac{\kappa N_S(N_S+1)}{(\kappa N_S + N_B+1)^2}\sum_{m=-(M-1)/2}^{(M-1)/2}{\rm Var}(|a_{R_m}|^2 \mid H_1) \nonumber \\[.05in]
&=& MN_B/4 + \frac{2M\kappa N_S(N_S+1)}{\kappa N_S + N_B + 1} \approx MN_B/4,
\end{eqnarray}
again reproducing the results from Sec.~\ref{HetHom}.

\section{Comparison Between QI PC and QI Hetero-Homodyne Reception}  \label{AppB}

The sufficient statistic, $\ell_{PC}$, for Guha and Erkmen's PC receiver is the outcome of measuring the observable
\begin{equation}
\hat{\ell}_{PC} \equiv \sum_{m=(M-1)/2}^{(M-1)/2} {\rm Re}[(\sqrt{2}\,\hat{a}_{V_m}^\dagger + \hat{a}_{R_m})\hat{a}_{I_m}],
\end{equation}
where the $\{\hat{a}_{V_m}\}$ are the vacuum-state, signal-port inputs to the PC receiver's parametric amplifier that is used to conjugate the return modes.  In QI's $N_S \ll 1$, $\kappa \ll 1$, $N_B \gg 1$, $M \gg 1$ asymptotic regime, the PC receiver's minimum error-probability decision rule is
\begin{equation}
\ell_{PC} \begin{array}{c}
\mbox{\scriptsize say $H_1$}\\ \ge \\ < \\ \mbox{\scriptsize $H_0$} \end{array} M\sqrt{\kappa N_S}\,/2,
\end{equation}
to decide between target absence and presence. This receiver's performance matches that of our hetero-homodyne receiver.  However, outside of that regime, its sufficient statistic is noisier than that of the hetero-homodyne receiver.  In particular, both receivers' sufficient statistics have the same conditional means under $H_0$ and $H_1$, but
\begin{equation}
{\rm Var}(\ell_{PC} \mid H_0) = \frac{M}{4}[(N_B+1)(N_S+1) + N_BN_S + 2N_S]
\end{equation}
and
\begin{align}
{\rm Var}(\ell_{PC} \mid H_1) &= \frac{M}{4}[(N_R+1)(N_S+1) + N_RN_S \nonumber \\[.05in]
&\hspace{.2in}+\,\,2N_S+2\kappa N_S(N_S+1) ] 
\end{align}
where $N_R \equiv \kappa N_S+N_B$,
exceed their hetero-homodyne receiver counterparts from Appendix~\ref{AppA}. The reason for this behavior is that the PC receiver's phase-conjugation operation admits more quantum noise than does the hetero-homodyne receiver. 

\section{Hetero-Homodyne Detection with Idler Heterodyning \label{AppC}}
\begin{figure}[t!]
\centering
{\includegraphics[width=0.7\columnwidth]{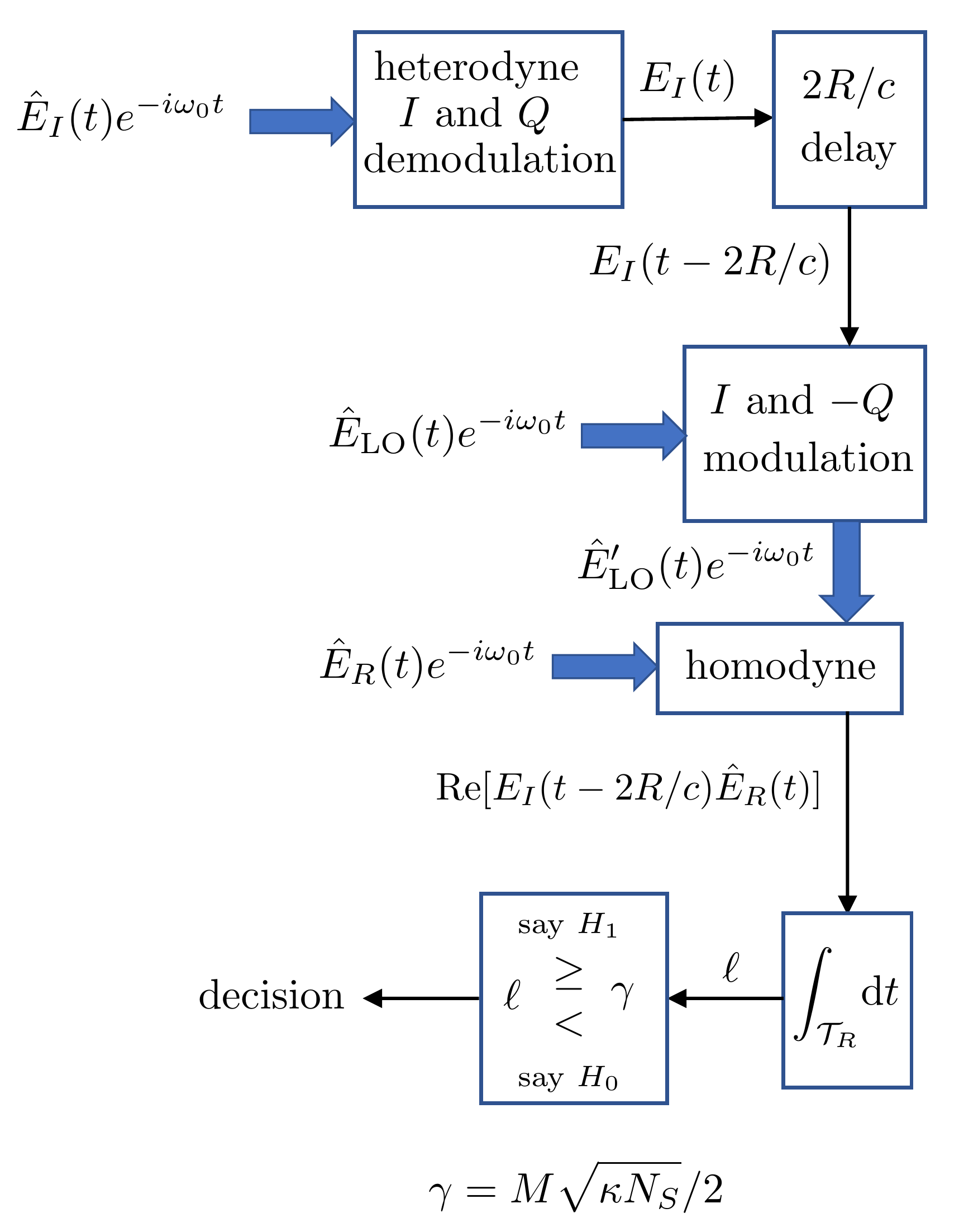}}
\caption{\label{HetHom2} Schematic of a hetero-homodyne receiver without quantum memory.  Thick arrows represent quantum microwave fields.  Thin arrows represent baseband signals that are conditioned on the outcome of the heterodyne measurement.}
\end{figure}

It is very tempting to replace the cascaded POVM shown in Fig.~\ref{HetHomFig} with the one shown in Fig.~\ref{HetHom2}, because it eliminates the need for idler storage in a quantum memory, as we now explain.

In this receiver architecture the roles of the returned radiation and the stored idler have been exchanged.  Heterodyne detection is now performed on the idler, yielding a classical random process $E_I(t)$ after $I$ and $Q$ demodulation.  This classical signal can be stored for the $2R/c$ range delay \emph{without} the need for a quantum memory.  Indeed, it can be sampled at the Nyquist rate for bandwidth $B$, quantized finely, and stored in a conventional digital memory for reconstitution when it is needed to drive the $I$ and $-Q$ modulation of the $\hat{E}_{\rm LO}(t)$ field.  The observable that is measured by homodyne detection, conditioned on $E_I(t-2R/c)$, is then $\hat{\ell} = {\rm Re}[E_I(t-2R/c)\hat{E}_R(t)]$.  Paralleling the development from Appendix~\ref{AppA}, it is easily shown that, with $\ell$ being the result of the $\hat{\ell}$ measurement, the following threshold test minimizes the error probability for equally-likely target absence and presence when $N_S \ll 1 $, $\kappa \ll 1$, $N_B \gg 1$, and $M\gg 1$:
\begin{equation}
\ell \begin{array}{c}
\mbox{\scriptsize say $H_1$}\\ \ge \\ < \\ \mbox{\scriptsize $H_0$} \end{array} \frac{\sqrt{\kappa N_S}}{2}\int_{\mathcal{T}_R}\!{\rm d}t\,|E_I(t-2R/c)|^2 \approx M\sqrt{\kappa N_S}/2.
\end{equation}
Again paralleling what we did in Appendix~\ref{AppA}, we find that
\begin{equation}
\mathbb{E}(\ell \mid H_0) = 0,
\end{equation}
\begin{equation}
\mathbb{E}(\ell \mid H_1) \approx M\sqrt{\kappa N_S},
\end{equation}
\begin{equation}
{\rm Var}(\ell \mid H_1) \approx {\rm Var}(\ell \mid H_0) = MN_B/2,
\end{equation}
for the Fig.~\ref{HetHom2} receiver in QI's asymptotic regime.  Invoking $M \gg 1$ to approximate $\ell$ as conditionally Gaussian given the true hypothesis, we get
\begin{equation}
\Pr(e)_{QI} = Q(\sqrt{M\kappa N_S/2N_B}) \le e^{-M\kappa N_S/4N_B}/2,
\end{equation}
which matches CI's error probability.  

Sadly, we have found that the Fig.~\ref{HetHom2} receiver offers no quantum advantage, i.e., the need for quantum memory cannot be circumvented in this manner.  The reason for the performance disparity between the Fig.~\ref{HetHomFig} and Fig.~\ref{HetHom2} forms of hetero-homodyne reception is easily seen.  Conditioned on $E_R(t)$, the idler is quantum limited, hence its homodyne detection suffers half the quantum noise of its heterodyne detection.  Conversely, conditioned on $E_I(t-2R/c)$, the returned radiation is background limited, hence its homodyne detection suffers the same noise as its heterodyne detection.


\begin{thebibliography}{99}
\bibitem{footnote0}Angle brackets are used here to denote quantum average.  They will also be used later to denote classical average.
\bibitem{footnote3}The coherent state $|E_S(t)\rangle_S$ satisfies $\hat{E}_S(t)|E_S(t)\rangle_S = E_S(t)|E_S(t)\rangle_S$ for $t\in \mathcal{T}_0$.  Note that all coherent states are single-mode states.  So, although all time-limited $E_S(t)$'s satisfying $\int\!{\rm d}t\,|E_S(t)|^2 = N_T$ give the same performance, to conform with the high time-bandwidth product that will necessarily be assumed for QI, we can regard $E_S(t)$ as a duration-$T$, bandwidth-$B$ chirped pulse with the same $M$ as QI.
\bibitem{footnote4}This modulator must be attenuating in order to avoid accruing quantum noise associated with amplification.
\end{thebibliography}
\end{document}